\documentclass[conference]{IEEEtran}
\IEEEoverridecommandlockouts
\usepackage{blkarray}                                      % to support matrix
\usepackage{algpseudocode}                                 % new algorithm package
\usepackage{algorithm}
\usepackage{graphicx}                                      % include pdf figures
\usepackage{amsmath}
\usepackage{amssymb}
\usepackage{amsfonts}
\usepackage{amsthm}
\usepackage[mathcal]{eucal}
\usepackage{mathrsfs}
\usepackage{booktabs}
\usepackage{enumerate}
\usepackage{multirow}
\usepackage{color}
\usepackage{cite}                                          % more citations in one bracket
\usepackage{comment}                                       % use comment
\usepackage{soul}                                          % use highlight command \hl{}
\soulregister\cite7
\soulregister\ref7
\soulregister\pageref7
\usepackage{etoolbox}                                      % commands \newtoggle, \toggletrue, \iftoggle
\usepackage{url}
\usepackage{nth}                                           % nth command
\usepackage{bm}                                            % bm command
\usepackage{courier}
\usepackage{balance}
\usepackage{threeparttable}
\usepackage{xcolor,colortbl}
\usepackage{footnote}
\usepackage{listings}
\usepackage{setspace}                                      % setstretch{}
\usepackage[inline]{enumitem}

\usepackage{verbatim}
\usepackage[bookmarks=false]{hyperref}
\hypersetup{
    colorlinks = true,
    citecolor  = blue,
    linkcolor  = blue,
    urlcolor   = blue,
}
\usepackage{tikz}
\usetikzlibrary{patterns,snakes}
\usetikzlibrary{positioning,calc,fit,decorations.pathmorphing,shapes.geometric, shapes.gates.logic.US, calc}
\usetikzlibrary{arrows,arrows.meta,decorations.markings,shapes,shapes.arrows}
\usetikzlibrary{decorations,decorations.pathreplacing}
\usetikzlibrary{backgrounds}
\usepackage{filecontents}                                  % support to pgfplots
\usepackage{pgfplots}
\usepackage{pgfplotstable}
\usepackage{scalefnt}
\pgfplotsset{compat=newest}
\usepackage{caption}
\usepackage{pifont}                                        % ding commend
\usepackage{cleveref}
\Crefformat{figure}{Fig.~#2#1#3}                           % "Fig.", instead of "Figure"
\Crefname{subfigure}{Fig.}{Figs.}
\Crefname{figure}{Fig.}{Figs.}
\Crefformat{table}{TABLE~#2#1#3}                           % "TABLE", instead of "Table"
\usepackage[figuresright]{rotating}
\definecolor{CUHKorange}{RGB}{244,106,18} %F47012
\definecolor{CUHKblue}{RGB}{0,111,190}    %006FBE
\definecolor{CUHKgreen}{RGB}{0,127,128}   %007F80
\definecolor{CUHKred}{RGB}{228,46,36}     %E42E24
\definecolor{CUHKyellow}{RGB}{198,148,34} %C69422
\definecolor{CUHKdark}{RGB}{114,44,114}   %722C72
\definecolor{CUHKmiddle}{RGB}{144,44,144} %902C90
\definecolor{CUHKlight}{RGB}{167,44,167} 
\definecolor{CUHKpurple}{RGB}{117,15,109}
\definecolor{CUHKgold}{RGB}{221,163,0}
\definecolor{CUHKribbon}{RGB}{244,223,176}
\definecolor{CUHKblack}{RGB}{34,24,21}

\renewcommand{\bf}[1]{\textbf{#1}}
\renewcommand{\tt}[1]{\texttt{#1}}
\renewcommand{\vec}[1]{\boldsymbol{#1}}    % re-define vec command

\newcommand{\minisection}[1]{\vspace{.06in}\noindent{\textbf{#1}}.}

\usepackage{tcolorbox}
\tcbuselibrary{skins,breakable}
    {\endtcolorbox}
    {\endtcolorbox}

\crefname{mytheorem}{Theorem}{Theorems}
\crefname{mylemma}{Lemma}{Lemmas}
\crefname{myclaim}{Claim}{Claims}
\crefname{myproperty}{Property}{Properties}
\crefname{mycorollary}{Corollary}{Corollaries}

\algrenewcommand\textproc{\texttt}

\makeatletter
\let\OldStatex\Statex
\renewcommand{\Statex}[1][3]{%
  \setlength\@tempdima{\algorithmicindent}%
  \OldStatex\hskip\dimexpr#1\@tempdima\relax
}
\makeatother

\RequirePackage[normalem]{ulem} %DIF PREAMBLE
\RequirePackage{color}\definecolor{RED}{rgb}{1,0,0}\definecolor{BLUE}{rgb}{0,0,1} %DIF PREAMBLE
\graphicspath{{./pic/}{../}}

\usepackage{textcomp}
\usepackage{xcolor}
\usepackage{comment}
\usepackage{booktabs}
\usepackage{multirow}
\usepackage{array}
\usepackage[labelformat=simple]{subcaption}

\usepackage{tabularx}

\usepackage[a4paper, total={184mm,239mm}]{geometry}
\def\BibTeX{{\rm B\kern-.05em{\sc i\kern-.025em b}\kern-.08em
    T\kern-.1667em\lower.7ex\hbox{E}\kern-.125emX}}

\usepackage{multirow}
\usepackage{xcntperchap}
\usepackage{tikz} 
\newcommand*\circled[1]{\tikz[baseline=(char.base)]{ \node[shape=circle,draw,inner sep=0.2pt] (char) {#1};}}

\newcommand{\tabitem}{~~\llap{\textbullet}~~}

\begin{document}

\title{\huge
    InF-ATPG: Intelligent FFR-Driven ATPG with Advanced Circuit Representation Guided Reinforcement Learning%\vspace{-0.008em}
    }
\author{
\IEEEauthorblockN{Bin Sun\textsuperscript{1,2}, Rengang Zhang\textsuperscript{1,2}, Zhiteng Chao\textsuperscript{1,2,3}, Zizhen Liu\IEEEauthorrefmark{1}\textsuperscript{1,2}, Jianan Mu\IEEEauthorrefmark{1}\textsuperscript{1,2},  
  Jing Ye\textsuperscript{1,2,3}, Huawei Li\IEEEauthorrefmark{1}\textsuperscript{1,2,3}}
  \IEEEauthorblockA{\textsuperscript{1}\textit{State Key Lab of Processors, Institute of Computing Technology, Chinese Academy of Sciences}\\
            \textsuperscript{2}\textit{University of Chinese Academy of Sciences} \\ 
                  \textsuperscript{3}\textit{CASTEST Co., Ltd.} \\ 
                \\  sunbin23@mails.ucas.ac.cn, 
                    zhangrengang23z@ict.ac.cn, \\
                  \{chaozhiteng20g,  
                    liuzizhen, mujianan, 
                    yejing, lihuawei\}@ict.ac.cn\\
                  }
}
\begin{comment}

{\footnotesize \textsuperscript{*}}
\thanks{Identify applicable funding agency here. If none, delete this.}
\author{\IEEEauthorblockN{1\textsuperscript{st} Given Name Surname}
\IEEEauthorblockA{\textit{dept. name of organization (of Aff.)} \\
\textit{name of organization (of Aff.)}\\
City, Country \\
email address or ORCID}
\and
\IEEEauthorblockN{2\textsuperscript{nd} Given Name Surname}
\IEEEauthorblockA{\textit{dept. name of organization (of Aff.)} \\
\textit{name of organization (of Aff.)}\\
City, Country \\
email address or ORCID}
\and
\IEEEauthorblockN{3\textsuperscript{rd} Given Name Surname}
\IEEEauthorblockA{\textit{dept. name of organization (of Aff.)} \\
\textit{name of organization (of Aff.)}\\
City, Country \\
email address or ORCID}
}
\end{comment}

\maketitle

\begin{abstract}
    % Automatic test pattern generation (ATPG) is essential for integrated circuit (IC) testing, tasked with producing test patterns to achieve high fault coverage. 
    % As semiconductor technology advances, traditional ATPG algorithms struggle with insufficient fault coverage and long execution times, necessitating optimization.
    Automatic test pattern generation (ATPG) is a crucial process in integrated circuit (IC) design and testing, responsible for efficiently generating test patterns.
    As semiconductor technology progresses, traditional ATPG struggles with long execution times to achieve the expected fault coverage, which impacts the time-to-market of chips.
    Recent machine learning techniques, like reinforcement learning (RL) and graph neural networks (GNNs), show promise but face issues such as reward delay in RL models and inadequate circuit representation in GNN-based methods. 
    In this paper, we propose InF-ATPG, an intelligent FFR-driven ATPG framework that overcomes these challenges by using advanced circuit representation to guide RL. By partitioning circuits into fanout-free regions (FFRs) and incorporating ATPG-specific features into a novel QGNN architecture, InF-ATPG enhances test pattern generation efficiency. 
    Experimental results show InF-ATPG reduces backtracks by 55.06\% on average compared to traditional methods and 38.31\% compared to the machine learning approach, while also improving fault coverage.
    %while improving fault coverage slightly. %by 0.12\%.
\end{abstract}

\section{Introduction}

Automatic test pattern generation (ATPG) is crucial for integrated circuit (IC) testing and design for test (DFT), tasked with generating efficient test patterns to achieve high fault coverage~\cite{jang1995design,barnhart2001opmisr}. 
As semiconductor technology advances into deep submicron processes, the complexity of chip design has increased significantly~\cite{lavagno2017electronic,kamal2022silicon},
making the generation of test patterns that meet fault coverage requirements increasingly time-consuming~\cite{zhen2023conflict}.
It is indicated that generating test patterns with sufficient fault coverage can take over 20\% of development time, often exceeding a month~\cite{wang2006vlsi}.
Therefore, optimizing ATPG algorithms has become a pressing challenge~\cite{kunz2002sat,bushnell2004essentials}.

\begin{figure}[tb!]
    \centering
    \subfloat[]{ \includegraphics[width=0.50\linewidth]{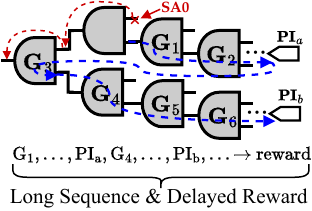} \label{fig:problem-a}}
    \subfloat[]{ \includegraphics[width=0.50\linewidth]{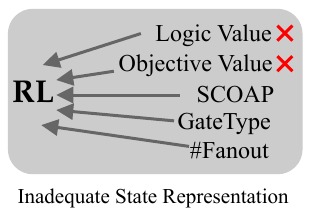} \label{fig:problem-b}}
    \caption{Challenges in existing methods: (a) Long decision sequences lead to delayed rewards; (b) Inadequate circuit state representation for ATPG.}
    \label{fig:problem}
\end{figure}

ATPG is a highly complex problem of finding signal assignments in the circuit space to satisfy specific constraints \cite{chen2013two}.
While traditional methods rely on heuristic algorithms, 
% recent advancements have introduced learning-based models aimed at enhancing searching capabilities through accumulated data rather than relying on ad-hoc optimizations based on human experience. 
recent advancements introduced learning-based models aimed at enhancing searching capabilities through accumulated data rather than relying on human-driven optimization. 
However, these intelligent models still face problems in decision-making and representation for ATPG tasks.
One major problem is delayed rewards: SmartATPG \cite{li2024smartatpg} uses reinforcement learning (RL) to make decisions gate-by-gate, as illustrated in \Cref{fig:problem-a}.
% As circuit depth increases, the decision chain of this gate-by-gate approach becomes excessively long, leading to reward delay and slow convergence. 
As the bandwidth and depth of the circuit increase, the fault support regions expand considerably, resulting in excessively long decision chains for gate-by-gate approaches. This elongation leads to delayed rewards and slow convergence in the optimization process.
Another key problem is the inadequate representation of decision values, where existing approaches like FGNN2 \cite{wang2024fgnn2} and DeepGate2 \cite{shi2023deepgate2} leverage graph neural networks (GNNs) to model circuit functionality and structure but fail to capture ATPG-specific features like logic states and SCOAP controllability/observability~\cite{goldstein1979controllability}. These logic states play a crucial role in ATPG algorithm decision-making, as illustrated in \Cref{fig:problem-b}. 
Thus, the main challenge for intelligent ATPG is how to efficiently represent circuits, including appropriate partitioning and state representation, to optimize RL decision-making and ensure critical ATPG-specific information is captured.

To this end, we propose InF-ATPG, an intelligent RL-based framework designed for efficient decision-making and representation in ATPG tasks. To enhance model performance, we extract broader circuit information by partitioning circuits into fanout-free regions (FFRs)~\cite{cong1994acyclic}, which reduces decision sequence length and captures higher-level structural insights. This abstraction integrates ATPG-specific features like logic states and SCOAP controllability/observability~\cite{goldstein1979controllability}, providing a more accurate representation of circuit states. Since simple GNNs struggle to fully capture the complexity of logic states, we introduce the Quality-Value Graph Neural Network (QGNN), which uses a logic-value-aware multi-aggregator to model complex relationships between circuit structure and functionality. Integrating QGNN into the RL agent enhances behavior-value estimation, enabling more efficient test pattern generation by offering deeper insight into circuit behavior. In summary, the paper makes three key contributions as follows:

\begin{itemize}
% [leftmargin=*]
	% \setlength{\itemsep}{0pt}
	% \setlength{\parsep}{0pt}
	% \setlength{\parskip}{0pt}

    % \item We identify FFR partitioning as the fundamental building block for an intelligent ATPG decision model. FFR-based partitioning shortens the decision chain and derives ATPG-specific logic states.
    % \item We propose an improved GNN framework that leverages a logic-value-aware multi-aggregator to learn ATPG-specific logic states, thereby optimizing the action-value estimation in the RL model.

    \item InF-ATPG innovatively explores block-level circuit spaces for RL-based ATPG. To address reward delay from long decision sequences, circuits are modeled with FFRs as the fundamental units, significantly shortening the sequence length. 
    \item To optimize the circuit representation and tailor the RL agent for ATPG-specific tasks, InF-ATPG integrates key circuit states and features into the QGNN framework. By capturing structural and functional features more efficiently at the FFR level, QGNN drives RL agents to achieve superior training and inference performance through a logic-value-aware multi-aggregator design.
    \item Experimental results show that InF-ATPG reduces backtracking by an average of 55.06\% compared to the traditional gate-level ATPG method, and by 38.31\% compared to the existing machine learning approach. Additionally, it reduces the decision sequence length by a factor of 7.0.
\end{itemize}

\section{Background and Related Work}
\label{sec:prelim}

% \begin{figure*}[tb!]
%     \centering
%     \includegraphics[width=0.92\linewidth]{pic/deepcell_overview.pdf}
%     \caption{
%         Overview of the InF-ATPG framework.
%         (a) Circuit partitioning into FFRs simplifies decision-making;
%         (b) QGNN-based state representation improves the RL agent’s ability to generalize;
%         (c) MDP and DQN modeling optimize test generation efficiency.
%     }
%     \label{fig:inf-atpg}
% \end{figure*}

% \begin{figure*}[tb!]
%     \centering
%     \includegraphics[width=0.92\linewidth]{pic/deepcell_arc.pdf}
%     \caption{
%         Overview of the InF-ATPG framework.
%         (a) Circuit partitioning into FFRs simplifies decision-making;
%         (b) QGNN-based state representation improves the RL agent’s ability to generalize;
%         (c) MDP and DQN modeling optimize test generation efficiency.
%     }
%     \label{fig:inf-atpg}
% \end{figure*}

\begin{figure*}[tb!]
    \centering
    \includegraphics[width=1.0\linewidth]{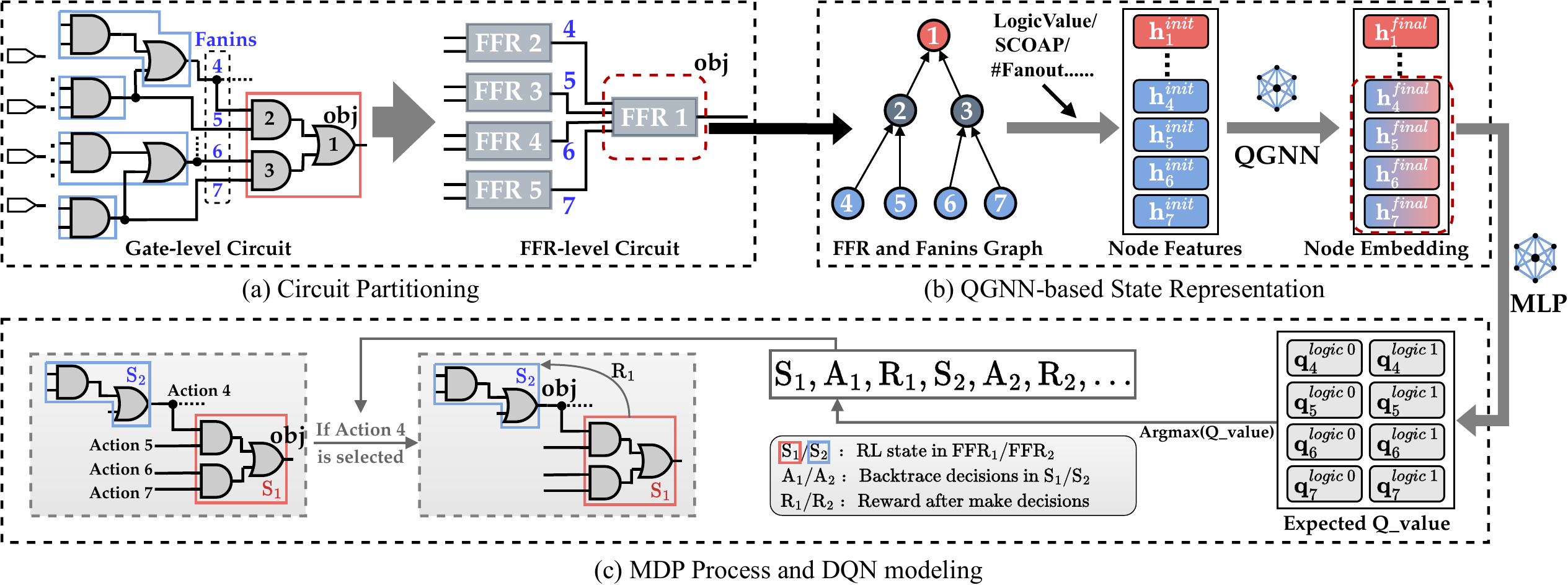}
    \caption{
        Overview of the InF-ATPG framework.
        (a) Circuit partitioning into FFRs simplifies decision-making;
        (b) QGNN-based state representation improves the RL agent’s ability to generalize;
        (c) MDP and DQN modeling optimize test generation efficiency.
    }
    \label{fig:inf-atpg}
\end{figure*}

As the complexity of IC design increases, traditional ATPG methods face numerous challenges in handling complex circuits.
Researchers have proposed various solutions to address these challenges. This section briefly reviews traditional ATPG methods, previous AI-guided methods, and their applications in ATPG.

\subsection{Preliminary ATPG Strategies and Fanout-Free Region}

\minisection{ATPG}
Traditional ATPG algorithms use heuristic searches to generate test patterns for fault coverage.
% Roth's D-algorithm \cite{b4} pioneered ATPG, marking the beginning of the field. 
Roth proposed D-algorithm\cite{b4} for conducting searches on circuits.
Goel introduced the PODEM \cite{b5} algorithm with distance-from-primary-inputs heuristic, which significantly reduced the decision space. At the same time, Fujiwara and Shimono proposed the FAN \cite{b6} algorithm, which avoids conflicts through unique sensitization and multiple backtracing. Larrabee further improved test generation for complex circuits by introducing SAT-based solving techniques \cite{larrabee1992test, huang2022neural, drechsler2008acceleration}. 
Building on this, researchers subsequently introduced heuristic optimization methods such as Simulated Annealing \cite{bandyopadhyay1998simulated, corno1997saara}, but these traditional method still struggle with growing circuit complexity.

% \minisection{Circuit Partitioning}
%Fanout Free Regions (FFRs) have been widely applied as a circuit partitioning method. By dividing circuits into multiple Fanout Free sub-blocks, FFR simplifies the circuit model \cite{cong1994acyclic}.

\begin{comment}
\minisection{Block-Level Circuit}
Fanout-free regions (FFRs) are commonly used for efficient circuit partitioning. By decomposing a circuit into FFRs, each with signal paths confined to a single output, FFRs simplify circuit structure and facilitate optimization processes \cite{cong1994acyclic}. It reduces complexity by localizing signal dependencies, enabling efficient test generation.
\end{comment}

%\minisection{Block-Level Circuit Partitioning}
\minisection{Fanout-Free Region}
% To address the limitations of traditional approaches in handling complex circuits, partitioning methods like Fanout-Free Regions (FFRs) have been introduced.
Fanout-free regions (FFRs) decompose a circuit into sub-regions in which every gate, except for the FFR head, has a single fanout, 
% where each gate except fanout has only one output \cite{cong1994acyclic}, 
simplifying the internal dependencies of each block. This localized structure reduces decision complexity, allowing more efficient test generation by confining the search space within smaller, independent regions. It should be noted that the FAN \cite{b6} algorithm utilizes FFR. FAN avoids conflicts through reverse implications across multiple gates, making decisions at FFR roots through a voting mechanism. However, when the number of gates capable of reverse implication in the circuit is insufficient, performance degradation occurs.
% where all signal paths converge to a single output

% To overcome these challenges, partitioning techniques like Fanout-Free Regions (FFRs) have been adopted. FFRs divide a circuit into blocks where all signal paths converge to a single output \cite{cong1994acyclic}, simplifying circuit dependencies. This approach reduces the decision complexity by localizing signal interactions within each block, enabling more efficient test generation by narrowing the search space to smaller, independent regions.

\subsection{Machine Learning Techniques}

\minisection{Reinforcement Learning}
Reinforcement learning (RL) learns optimal policies through interaction with the environment, making it particularly suitable for tasks requiring sequential decision-making. Sutton and Barto laid the theoretical foundation for RL \cite{sutton1998reinforcement}, while Watkins and Dayan's Q-learning algorithm provided a classical approach to action selection in RL \cite{watkins1992q}. Recently, deep reinforcement learning (DRL) has combined deep neural networks with RL models, such as Mnih $et\ al.$'s deep Q-network (DQN) \cite{mnih2013playing}. Schaul $et\ al.$ introduced Prioritized Experience Replay (PER), %further 
improving DQN's training efficiency and convergence speed \cite{schaul2015prioritized}.

\minisection{Graph Neural Networks}
Graph neural networks (GNNs) aggregate node characteristics through message-passing mechanisms and have gained widespread application in graph processing tasks \cite{niepert2016learning}. GraphSAGE introduced inductive representation learning to enhance GNN performance on large-scale graphs \cite{hamilton2017inductive}. Veličković $et\ al.$ proposed the graph attention network (GAT), which employs attention mechanisms for more effective feature aggregation in graphs with complex interactions \cite{velickovic2017graph}.

\minisection{Representation Learning for Circuits}
Representation learning enables models to capture the complex relationships between logic elements in circuits. FGNN and FGNN2 employed pre-training on circuit logic to significantly improve representation accuracy \cite{wang2022functionality, wang2024fgnn2}. DeepGate series also incorporated functional characteristics to generalize to complex circuits \cite{li2022deepgate, shi2023deepgate2, shi2024deepgate3}. The circuit transformer maintained circuit equivalence to generate high-quality circuit representations \cite{licircuit}, highlighting the potential of circuit representation learning. However, these methods often overlook critical ATPG-specific circuit states and fail to incorporate key features from %traditional 
ATPG, limiting their direct application to ATPG problems.

\subsection{Machine Learning Applications in ATPG}

Recently, machine learning techniques, particularly the combination of deep learning and RL, have offered new solutions for ATPG. Roy $et\ al.$ proposed artificial neural networks to guide the decision-making process in ATPG\cite{b12}. SmartATPG used RL to automate test pattern generation \cite{li2024smartatpg}. HybMT proposed a machine learning algorithm based on a hybrid meta-predictor, which significantly improved test generation efficiency \cite{sarangihybmt}. The application of reinforcement learning in ATPG has further demonstrated machine learning's potential in improving test generation efficiency \cite{li2024smartatpg, li2023intelligent}.

However, as circuit size increases, these AI-guided methods still face challenges. Major issues include reward delay caused by long decision sequences and inadequate circuit state modeling, making RL model training and convergence more difficult and limiting their practical effectiveness in ATPG.

\section{InF-ATPG Framework}
\label{sec:algo}

%As the complexity of integrated circuit design increases, traditional machine learning-assisted ATPG algorithms face challenges such as long decision sequences, reward delay, and inadequate circuit state representation. These challenges hinder fault coverage and execution efficiency, making it difficult to meet the demands of modern chip designs. To address these issues, the InF-ATPG framework proposes an improved circuit partitioning strategy based on Fanout Free Regions (FFRs) and incorporates a Quality-Value Graph Neural Network (QGNN) agent with reinforcement learning (RL) to significantly improve test pattern generation efficiency and fault coverage.

To address the aforementioned challenges introduced by existing solutions, we propose the InF-ATPG framework, which employs a novel RL-based approach utilizing block-level circuit search spaces and ATPG-specific circuit representations. 
% By integrating an FFR-based partitioning strategy with a QGNN-driven agent, InF-ATPG significantly improves test generation efficiency and fault coverage.
\Cref{fig:inf-atpg} illustrates an overview of the InF-ATPG framework. Its key innovations are as follows:
% To address the challenges above, the InF-ATPG framework introduces a novel RL-based approach driven by block-level circuit search spaces and ATPG-specific circuit representations. By combining an FFR-based partitioning strategy with a QGNN-driven agent, it significantly enhances test generation efficiency and fault coverage.

%As IC design complexity grows, machine learning ATPG methods face challenges such as long decision sequences, reward delay, and inadequate state representation, limiting fault coverage and efficiency. To address these issues, the InF-ATPG framework introduces an FFR-based partitioning strategy combined with a Quality-Value Graph Neural Network (QGNN) driven agent and reinforcement learning, significantly improving test generation efficiency and fault coverage.

\begin{itemize}
    \item \textbf{Circuit partitioning and FFR backtrace (see \Cref{fig:inf-atpg}(a)):} %The circuit is divided into multiple FFRs, where each FFR serves as the smallest decision-making unit. This partitioning strategy simplifies and shortens the decision sequence, reducing redundant decisions and effectively alleviating reward delay in large-scale circuits. 
    % The circuit is partitioned into FFRs, each serving as the smallest decision-making unit. This reduces decision sequence length and mitigates reward delay by eliminating redundant decisions. Further elaboration on this aspect is provided in Section~\ref{3.1}.
    We first partition the circuit into Fanout-Free Regions (FFRs), which serve as the smallest decision-making units. This partitioning reduces the decision sequence length and mitigates reward delay by eliminating redundant decisions. Further details are provided in Section~\ref{3.1}.

    \item \textbf{QGNN-based state representation (see \Cref{fig:inf-atpg}(b)):}
    % A customized QGNN architecture is introduced, where the GNN serves as the RL agent. It represents the state of FFR-based sub-circuits, enhancing the generalization ability of the RL process. Further elaboration on this aspect is provided in Section~\ref{3.2}. 
    % We present a customized QGNN architecture, designed to function as the RL agent by representing the states of FFR-based sub-circuits. QGNN improves the RL process’s generalization capability by effectively capturing circuit-level features. Further details are discussed in Section~\ref{3.2}.
    While reducing decision length improves efficiency, it is essential to represent circuit states accurately to ensure correct decisions. To this end, we introduce a customized Quality-Value Graph Neural Network (QGNN) architecture, which captures critical features of FFR-based sub-circuits. QGNN enhances the RL agent's generalization ability by learning complex relationships between circuit elements. Further details are discussed in Section~\ref{3.2}.
    
    \item \textbf{MDP process and DQN modeling (see \Cref{fig:inf-atpg}(c)):} 
    % Based on FFR partitioning and QGNN state representation, %the InF-ATPG framework models the ATPG problem as a Markov Decision Process (MDP). 
    % the ATPG problem is modeled as a Markov decision process (MDP). 
    % Key circuit state features are selected, and a DQN method is employed to evaluate the action values of selecting different fanin gates of each FFR. The agent is trained to optimize ATPG by selecting the optimal actions in each FFR, thereby improving test efficiency. Further elaboration on this aspect is provided in Section~\ref{3.3}.
    Building on the FFR partitioning and QGNN representation, we model the ATPG task as a Markov decision process (MDP). Key circuit state features are extracted, and a Deep Q-Network (DQN) is used to evaluate the action values of selecting different fanin gates within each FFR. Further elaboration on this aspect is provided in Section~\ref{3.3}.
\end{itemize}

\subsection{Circuit Partitioning and FFR-Level Backtrace}\label{3.1}

Traditional ATPG algorithms encounter challenges such as long decision sequences and reward delay, particularly in large-scale circuits. To address these issues, the InF-ATPG framework adopts a partitioning strategy based on FFRs.

\begin{figure}[tb!]
    \centering
    \includegraphics[width=0.92\linewidth]{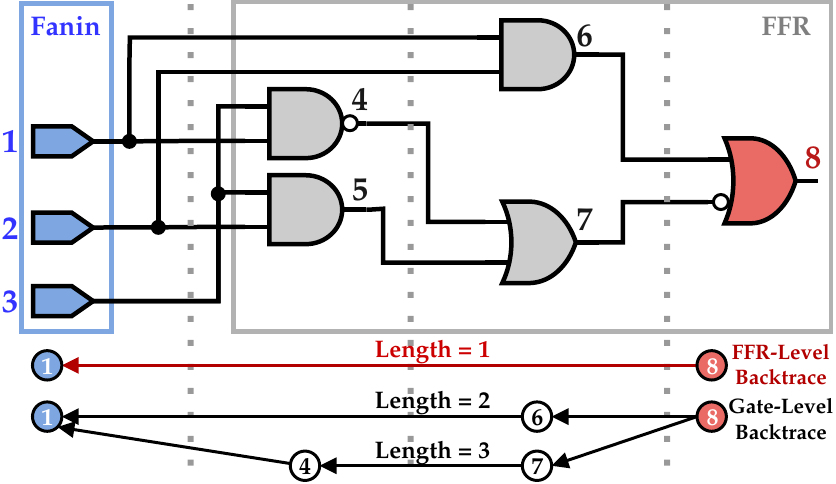}
    \caption{Comparison of FFR-level and Gate-level backtrace. FFR-level backtrace shortens the decision sequence by directly targeting fanin gates, while Gate-level backtrace involves a longer sequence of intermediate gates.}
    \label{fig:ffr_backtrace}
\end{figure}

An FFR, a circuit sub-block with non-overlapping internal logic and a single output, encapsulates a self-contained unit of logic. This structure simplifies the decision-making process while providing a higher-level abstraction of the circuit by isolating the logic within the FFR.
% An FFR is a circuit sub-block with non-overlapping internal logic and a single output node. Cross-layer circuit partitioning can shorten the decision sequence, potentially revealing broader structural information beyond the gate level. This structure simplifies the decision-making process by isolating logic within the FFR, allowing for more efficient test pattern generation.
As shown in \Cref{fig:ffr_backtrace}, FFR-level backtrace significantly reduces the decision sequence length compared to gate-level backtrace. The gate-level backtrace from the output gate \textit{8} to the fanin gate \textit{1} requires traversing multiple gates (i.e., gate \textit{6}, or gates \textit{7} to \textit{4}), resulting in a longer decision sequence (length 2 or 3). In contrast, FFR-level backtrace directly targets the fanin gates of the FFR, reducing the sequence to a length of 1 by bypassing intermediate gates and directly focusing on the higher-level structure of the circuit.

%By partitioning the circuit into FFRs, InF-ATPG reduces decision-making complexity and addresses reward delay. 
By extracting each FFR in the circuit and its associated fanin gates to form a sub-circuit, InF-ATPG reduces the complexity of decision-making and solves the problem of reward delay.
Each FFR becomes the smallest decision unit, allowing the framework to generate test patterns more efficiently than traditional gate-level approaches.

\subsection{QGNN-based State Representation}\label{3.2}

In RL, accurately representing the logic state of circuits is a key challenge, especially when ATPG problems are modeled as partially observable Markov decision processes (POMDPs). The precision of state modeling directly impacts the generalization ability of the RL model. Traditional ATPG methods primarily focus on individual gate characteristics, which limits their ability to capture complex circuit relationships, thereby affecting overall performance. To address this, the InF-ATPG framework introduces a customized quality-value graph neural network (QGNN) architecture. 
QGNN captures ATPG-specific features and introduces a logic-value-aware multi-aggregator design based on GAT\cite{velickovic2017graph}, enabling a more precise representation of the action's Q-value.  
% QGNN is specifically designed to capture the local fanout and reconvergent structures within circuits, providing enhanced state representation and improving the model's understanding of circuit logic.

\minisection{Circuit-Logic-State Features}
QGNN utilizes a variety of key circuit features to construct a comprehensive state representation, as shown in \Cref{tab:qgnn-features}.
The logic value directly reflects circuit functionality, while the objective value guides decision-making by adjusting targets. SCOAP controllability and observability quantify the difficulty of controlling and observing a node.
These features capture both the static and dynamic characteristics of circuit states that are essential for ATPG, enabling QGNN to model essential circuit logic states more effectively than heuristic methods. QGNN generates a more robust state representation, significantly enhancing the RL agent’s ability to make optimal decisions during ATPG.
\begin{table}[tb!]
    \centering
    \caption{Key circuit features in QGNN state representation}
    \resizebox{.99\linewidth}{!}{
    \begin{tabular}{|c|l|}
        \hline
        \textbf{Feature} & \textbf{Description} \\ \hline \hline
        Logic Value      & The logic value of the node (i.e., 0, 1, X) \\ \hline
        Objective Value  & Expected logic value for objective gate (in PODEM) \\ \hline
        SCOAP Controllability & Measures if the node can be controlled by PIs \\ \hline
        SCOAP Observability   & Measures if the node can be observed at a PO \\ \hline
        Gate Type        & One-hot encoded type of the logic gate \\ \hline
        \#Fanout         & Number of fanout gates connected to the node \\ \hline
        Depth            & Depth of the node from the PIs \\ \hline
    \end{tabular}
    }
    \label{tab:qgnn-features}
\end{table}
% The features capture both the static and dynamic circuit logic characteristics essential for ATPG.
% These features enable QGNN to model essential circuit logic states more effectively than heuristic methods, which often overlook such detailed properties. Considering factors such as logic value, controllability, and observability, QGNN generates a more robust state representation, significantly enhancing the RL agent’s ability to make optimal decisions during ATPG.

\begin{figure}[tb!]
    \centering
    \includegraphics[width=0.92\linewidth]{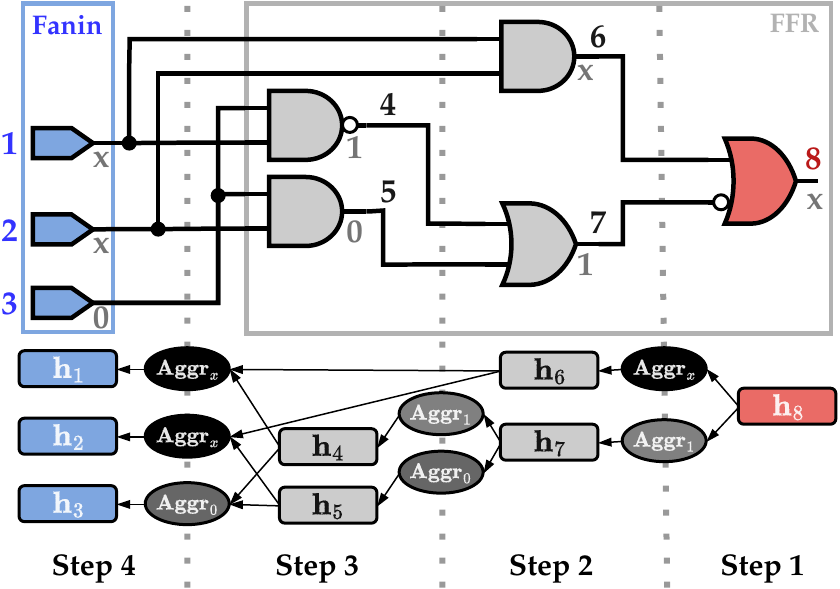}
    \caption{QGNN aggregation process across different logic states in an FFR. Nodes with 0, 1, and X logic values use different aggregators to propagate information, enhancing the model's ability to capture complex circuit dependencies.}
    \label{fig:qgnn_model}
\end{figure}

\minisection{QGNN Model}
QGNN models each FFR and its fanin gates as a graph, where each logic gate is represented as a node, and edges represent the connections between gates. As illustrated in \Cref{fig:qgnn_model}, the model assigns an embedding vector (e.g., $h_1$, $h_2$, $h_3$) to each node, encoding its features. These embeddings are then updated through message aggregating and passing using a GAT, which aggregates information in the reverse direction of circuit computation, layer by layer.
The aggregation process is governed by the following equations:
\begin{equation}
\begin{split}
    \alpha_{ij} &= softmax_{j}((\vec{W}_q \vec{h}_i)^\top \cdot (\vec{W}_k \vec{h}_j)), \\
    \vec{m}_j &= \vec{W}_v \vec{h}_j  \label{eq:alpha_ij}, \\
    \vec{msg}_i &= aggr(\vec{h}_j \mid j \in P(i)) = \sum_{j \in P(i)} (\alpha_{ij} \cdot \vec{m}_j), %\label{eq:msg_i}
\end{split}
\end{equation}
where $P(i)$ represents the predecessors of node $i$, $\alpha_{ij}$ represents the attention weight between node $i$ and node $j$, and $\vec{W}_q$, $\vec{W}_k$, $\vec{W}_v$ are the weight matrices of the aggregator. The aggregated message $\vec{msg}_i$ is then passed to a gated recurrent unit (GRU) to update the node's embedding:
\begin{equation}
    \vec{h}_i \leftarrow{ \text{GRU}(\vec{h}_i, aggr(\vec{h}_j \mid j \in P(i)))}.
\end{equation}

This process allows QGNN to capture the complex dependencies between nodes in the circuit, ultimately producing final embeddings for the target FFR. These embeddings provide rich information for the RL agent to select optimal actions during ATPG.

\minisection{Innovative Aggregator Design}
One of the key innovations in QGNN is the use of multiple aggregators tailored to different logic states ($0$, $1$, $X$).
As shown in \Cref{fig:qgnn_model}, different aggregators ($\text{aggr}_0$, $\text{aggr}_1$, $\text{aggr}_X$) are applied depending on the logic value of the node.
This design ensures that the information flow is adapted to the specific logic state of each node, allowing for more precise information propagation.
The aggregation functions for different logic states are defined as follows:
\begin{equation}
\begin{split}
\text{0-aggr} &\rightarrow \text{GRU}_{logic0}(\vec{h}_i, aggr_{logic0}(\vec{h}_j \mid j \in P(i))), \\
\text{1-aggr} &\rightarrow \text{GRU}_{logic1}(\vec{h}_i, aggr_{logic1}(\vec{h}_j \mid j \in P(i))), \\
\text{X-aggr} &\rightarrow \text{GRU}_{logicX}(\vec{h}_i, aggr_{logicX}(\vec{h}_j \mid j \in P(i))). \\
\end{split}
\end{equation}

This design is critical because nodes with different logic values (i.e., 0, 1, X) often exhibit different signal propagation behaviors. By decoupling the aggregation processes for different logic states, QGNN ensures that the information relevant to each state is processed independently, improving the expressiveness of the model.

\subsection{MDP and DQN Modeling}\label{3.3}
Building on FFR partitioning and QGNN-based state representation, the InF-ATPG framework models the ATPG problem as an MDP and incorporates a tailored DQN framework to train the QGNN-driven RL agent for optimal action selection.

\minisection{MDP Modeling}
To enable the RL agent to navigate the complexities of circuit testing effectively, the key elements of the MDP include:
\begin{itemize}
    \item \textbf{State:} The state is defined by the logic state and circuit features of each FFR. The QGNN generates embedding vectors that represent the state of each FFR in the circuit.
    \item \textbf{Action:} Based on the QGNN state embeddings, the RL agent selects actions, which involve choosing a fanin gate in each FFR for the backtrace process. The action space is unified across FFRs regardless of the number of fanins.
    \item \textbf{Reward:} The reward function is designed to reduce backtracks and prioritize critical primary inputs (PIs), thereby accelerating test generation. 
    \circled{1}The reward function penalizes unnecessary decisions with a small negative reward (-0.1). 
    \circled{2}When reaching a PI, the reward function assigns rewards based on the frequency of this PI's visits and the number of backtracks encountered between assigning a value to this PI and the next backtrace step. By this, we impose immediate penalties on conflicting PIs and avoid redundant visits. 
    % assigns penalties or rewards based on the number of backtracks and visits to specific PIs. 
    \circled{3}If a test is successfully generated or the backtrack limit is exceeded, the episode ends with a corresponding reward, thereby aligning the training process with the objective of test generation:
\end{itemize}
\[
%\centering
\text{Reward} =
\begin{cases} 
-0.1, & \text{if not reach PI}; \\
10 - \lambda_1 \cdot e^{\lambda_2 \cdot N}, & \text{if reach PI}; \\
100, & \text{Test found/No test}; \\
-100, & \text{Over backtrack limit};
\end{cases}
\]
\[
 N = \#\text{backtracks}[\text{PI}] + \#\text{visits}[\text{PI}].
\]

\begin{figure}[tb!]
    \centering
    \includegraphics[width=0.92\linewidth]{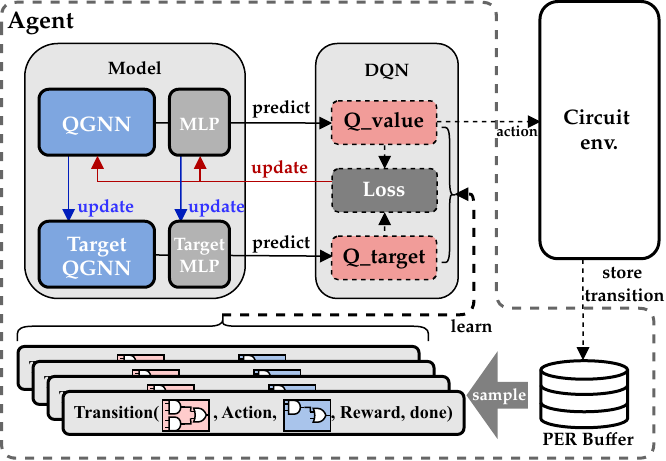}
    \caption{MDP and DQN process in InF-ATPG. The QGNN generates state embeddings, and the DQN evaluates action values, selects actions, and updates the model through reinforcement learning. Transitions are stored and sampled from a PER buffer to optimize learning efficiency.}
    \label{fig:dqn_model}
\end{figure}

\minisection{DQN Modeling}
The DQN process is illustrated in \Cref{fig:dqn_model}, and the key steps are as follows:
\begin{itemize}
    \item \textbf{Action Value Evaluation:} To enhance the effectiveness of circuit representation in RL-driven test generation, InF-ATPG employs the constructed QGNN alongside a multilayer perceptron (MLP) as the value network to guide the RL agent. This value network predicts the future value of all possible actions, with the action carrying the highest Q-value selected as the optimal choice.
    \item \textbf{Q-Value Update:} The QGNN and MLP are updated to minimize the loss between the predicted Q-values and the Q-targets. The Q-targets are computed using the target QGNN and target MLP, which is periodically updated.
    \item \textbf{Exploration-Exploitation Trade-off:} During training, exploration selects actions randomly to ensure the model explores diverse possibilities, while exploitation chooses actions based on the highest Q-value to optimize decision-making. The exploration rate starts high to encourage broad exploration and is gradually reduced as training progresses, shifting the model towards exploitation for more refined, optimal strategies.
    % \item \textbf{Exploration-Exploitation Trade-off:} During training, exploration encourages the model to try diverse actions, while exploitation leverages the knowledge gained to optimize decision-making. The balance between exploration and exploitation is dynamically adjusted through the exploration rate, promoting effective learning and convergence to an optimal strategy.
    % \item \textbf{Exploration-Exploitation Trade-off:} During training, the DQN balances exploration and exploitation by adjusting the exploration rate. As the model learns, more exploration is encouraged, followed by gradual convergence to an optimal strategy.  
\end{itemize}
% Based on the MDP formulation, InF-ATPG employs a Deep Q-Network (DQN) to train the RL agent for optimal action selection. The DQN process is illustrated in \Cref{fig:dqn_model}, and the key steps are as follows:
% Based on the MDP formulation, InF-ATPG employs a Deep Q-Network (DQN) method to train the QGNN and Multilayer Perceptron (MLP) for optimal action selection. The DQN process is illustrated in \Cref{fig:dqn_model}, and the key steps are as follows:

% \begin{itemize}
%     \item \textbf{Action Value Evaluation:} 
%     % The DQN evaluates the action values (Q-values) based on the current QGNN-generated embeddings. The action with the highest Q-value is selected as the optimal action.
%     DQN evaluates the actions based on the current model-predicted Q-value. The action with the highest Q-value is selected as the optimal action.
%     \item \textbf{Q-Value Update:} 
%     % The Q-values are updated by minimizing the loss between the predicted Q-values and the target Q-values. The target Q-values are computed using the target QGNN, which is periodically updated from the main QGNN model.
%     The QGNN and MLP are updated to minimize the loss between the predicted Q-values and the Q-targets. The Q-targets are computed using the target QGNN and target MLP, which is periodically updated.
%     \item \textbf{Exploration-Exploitation Trade-off:} During training, the DQN balances exploration and exploitation by adjusting the exploration rate. As the model learns, more exploration is encouraged, followed by gradual convergence to an optimal strategy.
% \end{itemize}

The training process leverages a prioritized experience replay (PER) buffer, where transitions (state, action, reward, next state, done) are stored during interaction with the circuit environment. Over time, the DQN samples transition from the PER buffer to learn and update. The QGNN and MLP are continuously updated based on the loss, 
% new Q-values,
ensuring the RL agent improves its decision-making over time.

\section{Experimental Results}
\label{sec:result}

% To evaluate the effectiveness of the proposed InF-ATPG framework, we conducted comprehensive experiments on various benchmarks. 
% This section introduces the experimental setup, benchmark circuit selection, and detailed analysis of the results.
% To comprehensively evaluate the performance of the proposed InF-ATPG framework, we conducted systematic experiments on a variety of benchmark circuits. 
% This section details the experimental setup, benchmark circuit selection, and an in-depth analysis of the results, showcasing the advantages of InF-ATPG over traditional ATPG methods and unimproved machine learning-based techniques.

\subsection{Experimental Setup}

All experiments were performed on NVIDIA A100 GPU and Intel
Xeon Processor (Skylake, IBRS) @ 2.99GHz, running Ubuntu 20.04. The primary goal was to evaluate the performance of InF-ATPG in comparison to the traditional ATPG method and the unimproved GNN-based approach. 
% As baseline methods, we used a standard gate-level ATPG heuristic based on PODEM, an FFR-level heuristic also based on PODEM, and an FFR-level GNN approach without the QGNN enhancements introduced in our framework.
As baseline methods, we used a standard PODEM based on gate-level heuristic \cite{oATPG}, an improved PODEM based on FFR-level heuristic, and an FFR-level GNN approach without the QGNN enhancements introduced in our framework.

%Benchmark circuits were selected from the ISCAS-85 \cite{brglez1985neutral} and ISCAS-89\cite{brglez1989combinational} suites to cover a broad spectrum of circuit sizes and complexities. The smallest circuit, \textit{c432}, contains 607 gates, while the largest, \textit{s38417}, consists of 19,768 gates. %These benchmarks were chosen to assess the scalability and robustness of InF-ATPG across various real-world IC testing scenarios.
Benchmark circuits from the ISCAS-85 \cite{brglez1985neutral} and ISCAS-89 \cite{brglez1989combinational} suites were selected to cover a range of circuit sizes and complexities, from 607 gates (\textit{c432}) to 19,768 gates (\textit{s38417}). 
% These benchmarks evaluate InF-ATPG's scalability and robustness in real-world IC testing.
It is noteworthy that the benchmark circuits were synthesized from RTL code and fault injection was performed using the oATPG\cite{oATPG} fault injection algorithm (oATPG\cite{oATPG} inserts faults at the pins of each cell to comply with industrial-grade ATPG workflows). Consequently, these benchmarks differ from the standard ISCAS-85 and ISCAS-89 benchmarks in both circuits and faults.

For hyperparameters, the QGNN and MLP learning rate were set to 0.0001, with a discount factor $\gamma$ of 0.95. 
The capacity of the PER buffer was configured as 512$\times$1024. 
Through extensive experimental exploration and parameter sensitivity analysis, the reward function parameters $\lambda_1$ and $\lambda_2$ were ultimately determined to be 7.5 and 0.07.
To ensure fairness and consistency, all comparison methods were evaluated under identical experimental conditions. 

For training, we selected 100 challenging faults from each circuit and conducted 10 training epochs. The remaining faults were used as the test set, allowing us to rigorously evaluate the generalization and effectiveness of the InF-ATPG framework.
Due to differences in implementation languages—InF-ATPG being a Python/C++ hybrid and the open-source gate-level ATPG heuristic \cite{oATPG} implemented in C++—a direct comparison of CPU time would be unfair. 
% Therefore, we use the number of backtracks as an indirect metric to assess ATPG performance.
Therefore, when comparing InF-ATPG with other baselines, we exclusively use the number of backtracks as a proxy metric to evaluate ATPG performance. Conversely, we directly compare the CPU runtime improvements of heuristic-based FFR-level ATPG versus Gate-level ATPG. By examining both the reduction in the number of backtracks between InF-ATPG and other methods, and the CPU runtime reduction between FFR-level ATPG and Gate-level ATPG, we can demonstrate that InF-ATPG has potential to achieve superior results through AI methods based on FFR-level decision-making.

% Due to implementation language differences, with InF-ATPG using Python/C++ hybrid and the open-source gate-level ATPG heuristic\cite{oATPG} implemented in C++, direct CPU time comparison is unfair. Therefore, we use the number of backtracks as an indirect measure to reflect ATPG performance.

\subsection{Experimental Results and Analysis}

\begin{table*}[tb!]
\centering
\caption{%Performance Comparison of InF-ATPG with gate-level and FFR-level ATPG Approaches
Comprehensive Performance Comparison of InF-ATPG with Gate-level and FFR-level ATPG Approaches%: Backtracks, Backtrace Steps, and Fault Coverage
}
\label{tab:metrics}

\begin{subtable}{\textwidth}
\centering
  \resizebox{\textwidth}{!}{

\begin{tabular}{|lr|rr|rrrr|rrrr|rrrr|} % 12 columns
\hline
\multirow{2}{*}{Circuit} & \multirow{2}{*}{\#Gates} & \multicolumn{2}{c|}{{InF-ATPG}} & \multicolumn{4}{c|}{Gate-level~\cite{oATPG}} & \multicolumn{4}{c|}{FFR-level heuristic} & \multicolumn{4}{c|}{FFR-level GNN} \\
    & & {\#B-track} & {\#B-trace} 
    & \#B-track  & {Red1}(\%) & \#B-trace & {Red2}(\%) 
    & \#B-track  & {Red1}(\%) & \#B-trace & {Red2}(\%) 
    & \#B-track  & {Red1}(\%) & \#B-trace & {Red2}(\%) \\
\hline
\hline
% \midrule
c432 & 607 & \textbf{27017} & \textbf{39417} & 46422 & 41.80 & 50809 & 22.42 & 41458 & 34.83 & 46165 & 14.62 & 47447 & 43.06 & 51587 & 23.59 \\ 
c1908 & 1514 & \textbf{8414} & \textbf{10903} & 14617 & 42.44 & 22568 & 51.69 & 13058 & 35.56 & 21654 & 49.65 & 11372 & 26.01 & 19264 & 43.40 \\ 
c1355 & 1792 & \textbf{19519} & \textbf{35861} & 47679 & 59.06 & 63557 & 43.58 & 118609 & 83.54 & 133023 & 73.04 & 55033 & {64.53} & 68635 & 47.75 \\ 
c2670 & 2335 & 37277 & \textbf{41753} & 41222 & 9.57 & 60486 & 30.97 & 32979 & -13.03 & 46616 & 10.43 & \textbf{29301} & -27.22 & 44949 & 7.11 \\ 
c3540 & 3080 & \textbf{32374} & \textbf{40576} & 72348 & 55.25 & 83859 & 51.61 & 96705 & 66.52 & 107593 & 62.29 & 64676 & 49.94 & 73104 & 44.50 \\ 
s9234 & 3518 & \textbf{5406} & \textbf{19909} & 75294 & {92.82} & 90508 & 78.00 & 82638 & {93.46} & 100165 & 80.12 & 10623 & 49.11 & 24620 & 19.13 \\ 
s15850 & 7290 & \textbf{10295} & \textbf{10295} & 51085 & 79.85 & 69527 & {85.19} & 43268 & 76.21 & 61271 & {83.20} & 24861 & 58.59 & 42603 & {75.84} \\ 
s38417 & 19768 & \textbf{53341} & \textbf{79765} & 132197 & 59.65 & 185653 & 57.04 & 89792 & 40.59 & 136154 & 41.42 & 92706 & 42.46 & 146852 & 45.68 \\ 

\hline
\hline
{Average} & - & - & - & - & {55.06} & - & {52.56} & - & {52.21} & - & {51.85} & - & {38.31} & - & {38.38} \\

    %\midrule
    % \bottomrule
    \hline
    % \hline
  \end{tabular}
  }
  \end{subtable}

\vspace{8pt} % 控制两个表格之间的距离，可以根据需要调整

\begin{subtable}{\textwidth}
\centering
\resizebox{\textwidth}{!}{
\begin{tabular}{|lr|rr|rrrr|rrrr|rrrr|} % 12 columns
\hline
\multirow{2}{*}{Circuit} & \multirow{2}{*}{\#Faults} & \multicolumn{2}{c|}{{InF-ATPG}} & \multicolumn{4}{c|}{Gate-level~\cite{oATPG}} & \multicolumn{4}{c|}{FFR-level heuristic} & \multicolumn{4}{c|}{FFR-level GNN} \\
    & & {\#UD-Fault} & {UFP(\%)} 
    & \#UD-Fault  & {Diff.}(\#) & UFP(\%) & {Imp.(\%)} 
    & \#UD-Fault  & {Diff.}(\#) & UFP(\%) & {Imp.(\%)} 
    & \#UD-Fault  & {Diff.}(\#) & UFP(\%) & {Imp.(\%)} \\
\hline
\hline
% \midrule

c432 & 1740 & \textbf{20} & \textbf{1.15} & \textbf{20} & 0 & \textbf{1.15} & 0.0 & \textbf{20} & 0 & \textbf{1.15} & 0.0 & \textbf{20} & 0 & \textbf{1.15} & 0.0 \\ 
c1908 & 4284 & \textbf{9} & \textbf{0.21} & \textbf{9} & 0 & \textbf{0.21} & 0.0 & 11 & 2 & 0.26 & 18.18 & \textbf{9} & 0 & \textbf{0.21} & 0.0 \\ 
c1355 & 4932 & \textbf{3} & \textbf{0.06} & \textbf{3} & 0 & \textbf{0.06} & 0.0 & 41 & 38 & 0.83 & 92.68 & 4 & 1 & 0.08 & 25.00 \\ 
c2670 & 6250 & \textbf{107} & \textbf{1.71} & \textbf{107} & 0 & \textbf{1.71} & 0.0 & \textbf{107} & 0 & \textbf{1.71} & 0.0 & \textbf{107} & 0 & \textbf{1.71} & 0.0 \\ 
c3540 & 8710 & \textbf{63} & \textbf{0.72} & 66 & 3 & 0.76 & 4.54 & 72 & 9 & 0.83 & 12.50 & \textbf{63} & 0 & \textbf{0.72} & 0.0 \\ 
s9234 & 6396 & \textbf{4} & \textbf{0.06} & 39 & 35 & 0.61 & 89.74 & 42 & 38 & 0.66 & 90.48 & \textbf{4} & 0 & \textbf{0.06} & 0.0 \\ 
s15850 & 13294 & \textbf{4} & \textbf{0.03} & 50 & 46 & 0.38 & 92.00 & 46 & 42 & 0.35 & 91.30 & 7 & 3 & 0.05 & 42.86 \\ 
s38417 & 36786 & \textbf{16} & \textbf{0.04} & 36 & 20 & 0.10 & 55.55 & 25 & 9 & 0.07 & 36.00 & 27 & 11 & 0.07 & 40.74 \\ 

\hline
\hline

{Average} & - & - & {\textbf{0.50}} & - & {+13} & 0.62 & {30.23} & - & {+17} & 0.73 & {42.64} & - & {+2} & 0.51 & {13.58} \\
    %\midrule
    % \bottomrule
    \hline
  \end{tabular}
  }
\end{subtable}

\vspace{0.36pt} % 表格和表注之间的距离

  \begin{tablenotes}[flushleft]
      \raggedright
 %  [leftmargin=*]
	% \setlength{\itemsep}{0pt}
	% \setlength{\parsep}{0pt}
	% \setlength{\parskip}{0pt}
  % \begin{itemize}
 % [leftmargin=*]
	% \setlength{\itemsep}{0pt}
	% \setlength{\parsep}{0pt}
	% \setlength{\parskip}{0pt}
   % \footnotesize
  \item \tabitem {Gate-level~\cite{oATPG}}: Gate-level PODEM algorithm using a distance-from-primary-inputs heuristic.
  \item \tabitem {FFR-level heuristic}: FFR-level PODEM algorithm using a distance-from-primary-inputs heuristic.
  \item \tabitem {FFR-level GNN}: FFR-level ATPG approach that utilizes unimproved GNN (without QGNN), enhancing with RL.
  \item \tabitem {\#B-track}: The total number of backtracks performed during ATPG.
  \item \tabitem {\#B-trace}: The total backtrace steps during ATPG refer to the process of tracing from the objective gate to PI.
  % \item \tabitem {\#Dec-Fault}: The number of detected faults in this method. \   \   \  
 % \   \  {FC(\%)}: Fault coverage in this method.  
  % \item \tabitem {Red1/Red2 (\%)}: Percentage reduction in the number of backtracks ({Red1}) and backtrace steps ({Red2}) compared to this method. Positive values indicate that the InF-ATPG approach achieved fewer backtracks and backtrace steps.
  % \item \tabitem {Diff1 (\#)/Diff2 (\%)}: The difference in the number of detected faults ({Diff1}) and fault coverage ({Diff2}) compared to this method. Positive values indicate improved fault detection and coverage.

  \item \tabitem {\#UD-Fault}: %Number of undetected faults in the method.
  Undetected faults number in the method. \   \   \   \ 
   {UFP(\%)}: %Percentage of undetected faults %relative 
   Undetected faults percentage
 to total %number of 
 faults in the method.  

\item \tabitem {Red1/Red2 (\%)}: Percentage reduction in the number of backtracks ({Red1}) and backtrace steps ({Red2}) compared to this method. Positive values indicate that the InF-ATPG approach achieved fewer backtracks and backtrace steps.

\item \tabitem {Diff. (\#)/Imp. (\%)}: The difference in the number of undetected faults ({Diff.}) and the improvement in UFP ({Imp.}) compared to this method. Positive values indicate a reduction in undetected faults and an improvement in UFP.
     
  % \end{itemize}
\end{tablenotes}
\end{table*}

\begin{table}[tb!]
    \centering
    \caption{CPU time comparison between FFR-level and gate-level ATPG}
    \label{tab:cpu_rumtimes}
    \resizebox{\linewidth}{!}{
        \begin{tabular}{|c|c|c|c|}
            \hline
            {Circuit} & {FFR-level(ms)} & {Gate-level~\cite{oATPG}(ms)}& {CPU Time Red.(\%)} \\
            \hline \hline
            c432  & 221 & 117 & -88.89 \\ 
            c1908 & 112  & 91  & -23.07  \\ 
            c1355 & 705 & 243 & -190.12  \\ 
            c2670 & 221 & 209  & -5.74  \\ 
            c3540 & 1008 & 1028 & \textbf{1.95}  \\ 
            s9234 & 441  & 495 & \textbf{10.91} \\ 
            s15850 & 695  & 731  & \textbf{4.92}  \\ 
            s38417 & 1629 & 2667 & \textbf{39.20} \\ \hline 
            % \hline
            % Average & - & - & 7.0  \\
            % \hline
        \end{tabular}
        }
        \vspace{0.36pt} % 表格和表注之间的距离
        \begin{tablenotes}[flushleft]
            \raggedright
        \item \tabitem  {CPU Time Red.(\%)}: Reduction of CPU time by FFR-level ATPG, calculated as (Gate-level$-$FFR-level) / (Gate-level).

\end{tablenotes}
\end{table}

\begin{table}[tb!]
    \centering
    \caption{Decision sequence length comparison between InF-ATPG and gate-level ATPG}
    \label{tab:benchmark_circuits}
    \resizebox{\linewidth}{!}{
        \begin{tabular}{|c|c|c|c|c|}
            \hline
            {Circuit} & {FFR Avg-Dep.} & {InF-ATPG} & {Gate-level~\cite{oATPG}}& {Ratio ($\times$)} \\
            \hline \hline
            c432    & 5.0 & \textbf{119234} & 1169414 & 9.8  \\ 
            c1908   & 5.9 & \textbf{89363}  & 549736  & 6.2  \\ 
            c1355   & 8.2 & \textbf{335034} & 1844036 & 5.5  \\ 
            c2670   & 4.8 & \textbf{173617} & 700113  & 4.0  \\ 
            c3540   & 6.2 & \textbf{289443} & 2371960 & 8.2  \\ 
            s9234   & 3.7 & \textbf{70286}  & 1104367 & 15.7 \\ 
            s15850  & 3.5 & \textbf{70669}  & 645829  & 9.1  \\ 
            s38417  & 3.9 & \textbf{266244} & 1521569 & 5.7  \\ \hline \hline
            Average & 5.2 & \textbf{176736} & 1238378 & 7.0  \\

            \hline
        \end{tabular}
        }
        \vspace{0.36pt} % 表格和表注之间的距离
        \begin{tablenotes}[flushleft]
            \raggedright
        \item \tabitem  {FFR Avg-Dep.}: Average depth of FFRs in the circuit.
        %\item \tabitem  {InF-ATPG Dec-Len.}: Decision sequence length for InF-ATPG, %representing the total number of decisions made during ATPG. representing number of decisions during ATPG.
        %\item \tabitem  {Gate-level Dec-Len.}: Decision sequence length for Gate-level heuristic PODEM\cite{oATPG}, showing the number of decisions in a conventional approach.
        \item \tabitem  {Ratio ($\times$)}: Reduction of decision sequence length by InF-ATPG, calculated as (Gate-level) / (InF-ATPG).
            % \end{itemize}
\end{tablenotes}

\end{table}

\begin{figure}[tb!]
    \centering
    \includegraphics[width=0.92\linewidth]{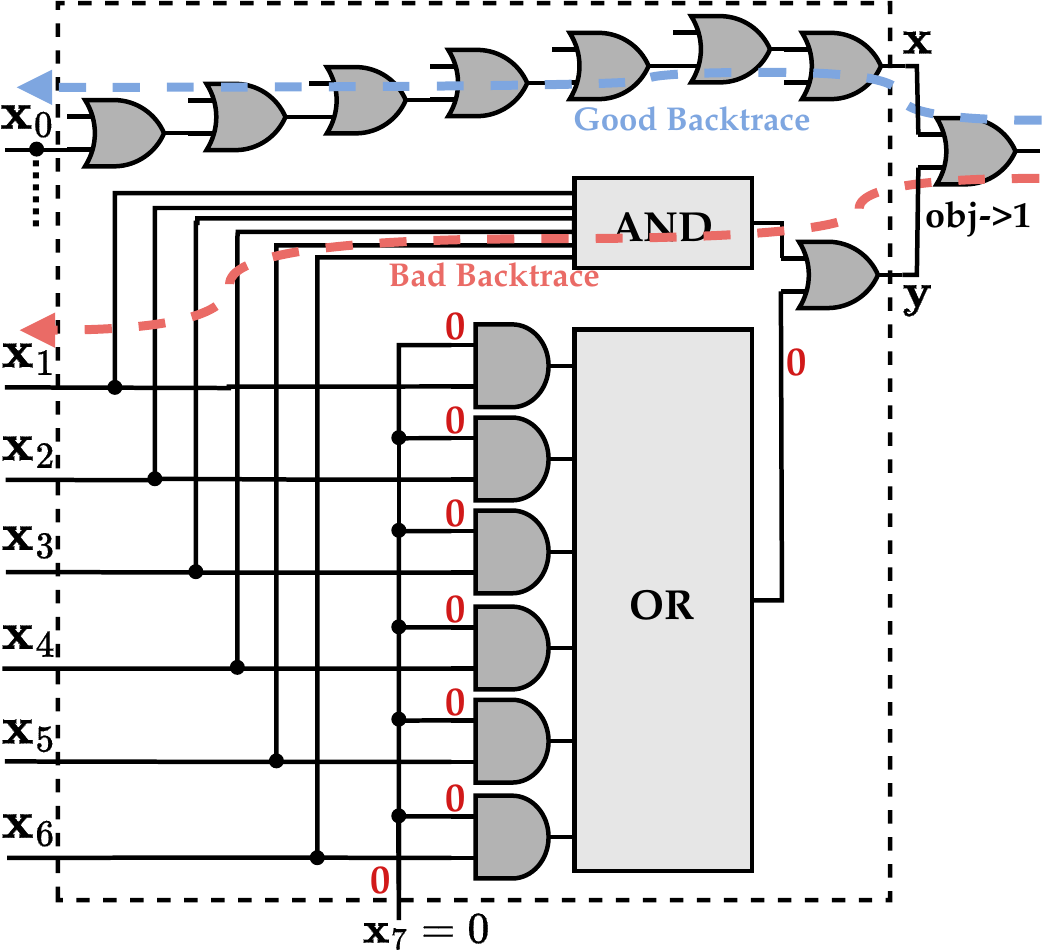}
    \caption{Gate-level decision-making, constrained by single-gate information, results in suboptimal backtrace decisions. In this instance, with \textit{x7} assigned to 0, setting gate \textit{y} to logic 1 would require all inputs \textit{x1}-\textit{x6} to be 1. Conversely, FFR-level decision-making can identify that a single backtrace through gate \textit{x} efficiently achieves the objective gate's target.}
    \label{fig:pic_5}
\end{figure}

This section presents a concise analysis of the experimental results focused on backtrack reduction and fault coverage improvement. 
Additionally, we conducted experiments to further validate our method's effectiveness by demonstrating the CPU runtime reduction of FFR-level compared to Gate-level approaches, as well as the decision sequence length reduction of InF-ATPG relative to Gate-level approaches.
% Additionally, we conducted experiments to further validate the effectiveness of our approach by demonstrating the reduction in decision sequence length. 
We compare InF-ATPG with gate-level and FFR-level ATPG approaches, and include an ablation study to highlight the contributions of key components.

\minisection{Backtrack and Backtrace Reduction}
%As shown in Table \ref{tab:metrics}, InF-ATPG achieves substantial reductions in backtracks and backtrace steps across benchmark circuits. 
As shown in the upper part of \Cref{tab:metrics}, InF-ATPG achieves a significant reduction in backtracks and backtrace steps in various circuits when controlling for different methods to achieve the same fault coverage.
% InF-ATPG achieves a significant reduction in backtracks and backtrace steps in various benchmark circuits when controlling for different methods to achieve the same fault coverage measurements backtrack and backtrace steps.
On average, InF-ATPG reduces backtracks by 55.06\% compared to gate-level heuristic ATPG\cite{oATPG} and by 52.21\% compared to FFR-level heuristic ATPG. Moreover, compared to FFR-level GNN without QGNN enhancements, InF-ATPG reduces backtracks by 38.31\% and backtrace steps by 38.38\%. The consistent improvements across different circuits underscore the robustness of the proposed method.

%The key factor driving these reductions is the FFR-based partitioning strategy, which effectively simplifies decision-making by breaking the circuit into smaller, independent sub-blocks. This reduces the length of decision sequences, thus minimizing the number of backtracks required to generate a complete set of test patterns. Moreover, the QGNN-enhanced state representation further aids the RL agent by providing more accurate information about circuit states, enabling it to make more informed decisions. Ultimately, these results demonstrate that InF-ATPG offers a significant reduction in computational complexity, which is especially crucial when dealing with large-scale circuits.

The FFR-based partitioning strategy simplifies decision-making by dividing circuits into smaller, independent subblocks, reducing decision sequence length and minimizing backtracks for ATPG. Additionally, the QGNN-enhanced state representation provides the RL agent with more accurate circuit information, enabling more informed decisions. This highlights InF-ATPG's ability to significantly reduce computational complexity, especially for large-scale circuits.

% \minisection{Fault Detection and Coverage}

%Table \ref{tab:metrics} also compares the fault detection capabilities and fault coverage (FC) of InF-ATPG with other methods. InF-ATPG consistently achieved high fault coverage across all circuits, with an average fault coverage of 99.5\%, outperforming both gate-level and FFR-level heuristic methods. The average fault coverage improvement compared to gate-level ATPG is +0.12\%, while the improvement over FFR-level heuristic methods is +0.2\%.

%The performance of InF-ATPG in fault detection can be attributed to the QGNN-based state representation, which captures key circuit features such as logic values, controllability, and observability. This allows the RL agent to identify and propagate faults more effectively, leading to higher fault coverage.

%%%

%InF-ATPG also excels in fault detection, achieving an average fault coverage of 99.5\% (Table \ref{tab:metrics}), outperforming both gate-level and FFR-level heuristics. The improvements in fault coverage—+0.12\% over gate-level ATPG and +0.2\% over FFR-level heuristics—are attributed to the QGNN-based state representation, which effectively captures circuit features like logic values, controllability, and observability, enabling more robust fault propagation and detection.
\minisection{Fault Detection and Coverage}
%InF-ATPG also demonstrates superior performance in fault detection, achieving an average fault coverage of 99.5\%, as shown in \Cref{tab:metrics}. 
% As shown in the lower part of \Cref{tab:metrics}, InF-ATPG also performs well in fault detection with an average fault coverage of 99.5\% when comparing fault coverage as in the same max-backtrack limit. 
% This represents a +0.12\% improvement over gate-level heuristic ATPG\cite{oATPG} and a +0.23\% improvement over FFR-level heuristic ATPG. These enhancements are largely due to the QGNN-based state representation, which captures crucial circuit features such as logic values, controllability, and observability. By leveraging this detailed state information, the RL agent is able to propagate faults more efficiently and generate more effective test patterns, leading to higher fault coverage.
As presented in the lower section of \Cref{tab:metrics}, under the same max-backtrack limit, 
InF-ATPG successfully detects hard-to-detect faults that other methods fail to detect. InF-ATPG achieves an average undetected fault percentage (UFP)
of 0.50\%. 
% InF-ATPG also performs well in fault detection, achieving an average undetected fault percentage (UFP) of 0.50\%.
% As presented in the lower section of \Cref{tab:metrics}, InF-ATPG also performs well in fault detection, achieving an average undetected fault percentage (UFP) of 0.50\% under the same max-backtrack limit. 
This represents a 30.32\% improvement compared to gate-level heuristic ATPG \cite{oATPG}, which has a UFP of 0.62\%, and a 42.64\% improvement over FFR-level heuristic ATPG, which has a UFP of 0.73\%.
% As shown in the lower part of \Cref{tab:metrics}, InF-ATPG also performs well in fault detection with an average undetected fault percentage (UFP) of 0.50\% when comparing fault coverage as in the same max-backtrack limit. 
% This marks a 30.32\% improvement over gate-level heuristic ATPG \cite{oATPG}, which has a UFP of 0.62\%, and a 42.64\% improvement over FFR-level heuristic ATPG, which has a UFP of 0.73\%. 

While the average improvement is notable, InF-ATPG shows particularly significant gains on larger circuits, such as \textit{s9234} and \textit{s15850}, where undetected faults were reduced by over 90\% compared to other methods.
These improvements are attributed to the QGNN-based state representation, which captures deeper circuit features, allowing RL agent to optimize test generation. 

\minisection{CPU Runtime}
As illustrated in \Cref{tab:cpu_rumtimes}, this experiment aims to demonstrate that although both approaches utilize distance-based heuristic methods, decision-making at the FFR-level yields superior results compared to Gate-level decision-making. Table 3 reveals that for larger-scale circuits, FFR-level decision-making contributes to a reduction in CPU runtime. For instance, in the \textit{s38417} circuit, FFR-level decision-making reduced runtime by 39.20\% compared to the Gate-level approach\cite{oATPG}. Conversely, \Cref{tab:cpu_rumtimes} indicates that for smaller-scale circuits, FFR-level decision-making incurs significant initialization overhead due to the computational requirements of FFR calculation, thus requiring longer CPU time compared to the Gate-level approach\cite{oATPG}.

FFR-level decision-making provides a broader perspective compared to Gate-level decision-making. As exemplified in \Cref{fig:pic_5}, to set the objective gate to logic 1, Gate-level decision-making tends to select gate \textit{y} with smaller logic depth. However, in this example, since \textit{x7} has already been assigned a value of 0, adjusting gate \textit{y} to 1 incurs a substantial cost (requiring \textit{x1}-\textit{x6} all to be set to 1). Consequently, selecting gate y for backtrace is not an optimal choice. In contrast, FFR-level decision-making has the opportunity to comprehensively consider the logical state and topological connectivity relationships at the subcircuit level, thereby recognizing that performing a backtrace through gate \textit{x} can achieve the objective gate's target more efficiently.

\minisection{Decision Sequence Length Reduction}
This experiment was designed to demonstrate the respective impacts of FFR and QGNN on reducing the decision sequence length.
\Cref{tab:benchmark_circuits} shows that InF-ATPG reduces decision sequence length by an average factor of 7.0 compared to gate-level ATPG\cite{oATPG}, with a notable reduction of 15.7 in the case of \textit{s9234}. 

This reduction stems from the FFR partitioning strategy, which simplifies the decision process by dividing the circuit into smaller, independent regions, each with an average FFR depth of 5.2. This coarse-grained approach reduces the complexity of decision sequences by focusing on broader, higher-level sub-circuits. 
% , rather than individual gates.
Also, %using FFRs obtains an average reduction of 7.0 which is larger than 5.2, enhancing decision quality by enabling the extraction of more comprehensive circuit information. 
InF-ATPG achieves an average reduction %in decision sequence length 
of 7.0, which surpasses the average FFR depth of 5.2. 
% This indicates an improvement in decision quality as it facilitates the extraction of more comprehensive circuit information.
Guided by QGNN’s enriched state representation, the RL agent leverages this broader context to make more informed decisions, reducing unnecessary steps, and further improving efficiency. This combination of coarse-grained partitioning and enhanced decision-making allows InF-ATPG to significantly shorten decision sequences, making it scalable for larger circuits.

The results confirm that InF-ATPG substantially reduces backtracks, enhances fault coverage, and shortens decision sequences. The ablation study underscores the critical role of FFR partitioning in reducing decision complexity, the effectiveness of QGNN-based state representation in capturing circuit logic for improved fault detection, and the contribution of RL in optimizing decision-making.

\section{Conclusion and Future Work}
\label{sec:conclu}

In this paper, we introduce InF-ATPG, a novel RL-based ATPG approach that leverages block-level circuit search spaces and advanced ATPG-specific circuit representations. By modeling circuits at the FFR level, we effectively reduce the decision sequence length, thereby addressing the reward delay issues commonly encountered in RL models. Furthermore, integrating key circuit features into the QGNN framework enhances the representation of circuit states and functionalities by decoupling the aggregation processes for different logic states. Experimental results demonstrate a 55.06\% reduction in backtracks compared to the traditional gate-level method and a 38.31\% reduction over the unimproved machine learning approach.

Future work could investigate the integration of more sophisticated reinforcement learning algorithms and additional circuit characteristics to further enhance the efficiency and accuracy of the ATPG process. Moreover, exploring efficient deployment strategies for these models within existing ATPG frameworks represents a compelling research direction that could significantly advance the practical applicability of AI-guided ATPG methods in industrial settings.

\newpage
\bibliographystyle{IEEEtran}
% \balance
\bibliography{ref}
% \printbibliography

\end{document}